# Robust Estimation of Resource Consumption for SQL Queries using Statistical Techniques

Jiexing Li
Department of Computer Sciences
University of Wisconsin - Madison
1210 West Dayton Street
Madison, WI 53706 USA
jxli@cs.wisc.edu

Arnd Christian König, Vivek Narasayya, Surajit Chaudhuri
Microsoft Research
Redmond, WA 98052, USA
{chrisko, viveknar, surajitc}@microsoft.com

## ABSTRACT

The ability to estimate resource consumption of SQL queries is crucial for a number of tasks in a database system such as admission control, query scheduling and costing during query optimization. Recent work has explored the use of statistical techniques for resource estimation in place of the manually constructed cost models used in query optimization. Such techniques, which require as training data examples of resource usage in queries, offer the promise of superior estimation accuracy since they can account for factors such as hardware characteristics of the system or bias in cardinality estimates. However, the proposed approaches lack robustness in that they do not generalize well to queries that are different from the training examples, resulting in significant estimation errors. Our approach aims to address this problem by combining knowledge of database query processing with statistical models. We model resource-usage at the level of individual operators, with different models and features for each operator type, and explicitly model the asymptotic behavior of each operator. This results in significantly better estimation accuracy and the ability to estimate resource usage of arbitrary plans, even when they are very different from the training instances. We validate our approach using various large scale real-life and benchmark workloads on Microsoft SQL Server.

## 1. INTRODUCTION

The ability to estimate resource consumption, such as CPU time, memory, logical page reads, etc., of SQL queries is crucial to various tasks in a database system. One such task is admission control: when a query is submitted into a DBMS, the system has to consider the resource requirements of the query as well as the available resources to determine if the query should be allowed to execute or be delayed. Another application is query optimization where estimates of resource usage are used to assign an overall cost to candidate execution plans.

Currently, resource estimation is based on manually constructed models, which are part of the query optimizer and typically use combinations of weighted estimates of the number of tuples flowing through operators, column widths, etc. Unfortunately, such models often fail to capture several factors that affect the actual resource consumption. For example, they may not include detailed modeling of all of the various improvements made to database query processing – such as *nested loop* optimizations [13, 11] which localize references in the inner subtree, and introduce "partially blocking" batch sorts on the outer side, thereby increasing the memory requirements and CPU time and reducing I/O compared to the traditional iterator model. Similarly, they may not accurately reflect specific hardware characteristics of the current production system or the impact of cardinality estimation errors.

To illustrate the issues with how resource-consumption is currently modeled, we executed a set of queries taken from the TPC-H [3] benchmark (executed on skewed data ($Z = 1$), on TPC-H datasets of scale-factors 1,2,4,6,8,10) and for each query measured the estimated CPU usage and the actual CPU time. To ensure that any errors in the estimates were mainly due to the way CPU time is currently modeled and not a result of errors in cardinality estimation, we only considered queries for which the cardinality estimates at each node in the executed query plan were within 90%-110% of the correct cardinalities – all other queries were removed. We plotted the results in Figure 1. We also plotted the linear regression line resulting from fitting these points using linear least-squares regression (which can be seen as an error-minimizing mapping of the optimizer-estimated cost (which is not measured in ms) to CPU-time). As we can see, even for the mapping chosen to minimize the overall error, there are significant differences between the estimated CPU cost and real CPU time for many queries. Similar observations have been made in other works as well (e.g., [15, 8]). Thus, although the optimizer models are very general and are applicable to any execution plan, they can incur significant errors.

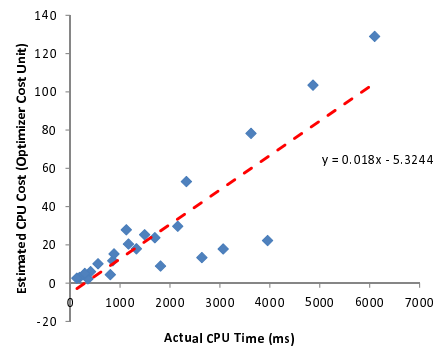

**Figure 1: Optimizer estimates can incur significant errors.**

Recent work has investigated the use of machine-learning based models for the estimation of resource usage (and run-time) of SQL queries [8, 12, 15, 7]. In these approaches, statistical models are trained on actual observations of the resource consumption of training queries, each of which is modeled using a set of features. When given sufficient training data, these statistical models can fit complicated resource-consumption curves more accurately than the hand-crafted functions used in query optimization. Because the models are trained on actual observations of resource consumption, they can capture a wide range of effects due to special cases in query processing, hardware architectures, and can compensate to some degree for systematic bias in cardinality estimation.

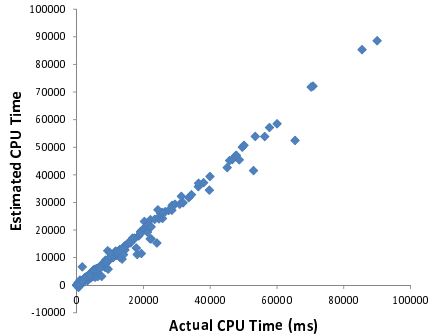

**Figure 2:** Statistical techniques can improve estimates significantly.

To illustrate the accuracy gains possible by using statistical techniques, consider Figure 2 which displays the estimates and actual CPU times for one of the experiments on TPC-H queries in this paper. In this experiment, we train a model using over 2500 TPC-H benchmark queries (generated using the QGEN tool over highly skewed data which ensures high variance in the resource consumption within a query template) and test it on a *disjoint* set of TPC-H benchmark queries (i.e., we use the same templates, but different parameters). As we can see, the estimated and actual CPU-times approximate a diagonal lines much more closely than in Figure 1 and there are no queries with significant errors.

## 1.1 The Need for Robust Models

While statistical techniques can improve estimation accuracy significantly, they can also fail dramatically when the queries they are used on differ significantly from the queries used to train the models. Examples of such differences may be changes in the size or distribution of the data accessed by a query, changes in its execution plan caused by statistics updates, or simply execution of a previously unseen "ad-hoc" query. For resource estimation, which is a key component of a DBMS, such lack of robustness would be unacceptable. We say a model is *robust* if differences in features between the test and training queries do not result in a significant degradation of the estimation accuracy.

Unfortunately, lack of robustness is a significant issue for statistical techniques that have been proposed previously for resource estimation of a query execution plan. For example, consider the KCCA model used in [15], which estimates resource consumption by *averaging* the resource consumption of *similar* queries in the training data. Therefore, if such a model is used for a query whose resource-usage is much larger than all training queries, the resulting estimate will necessarily be too low and potentially "off" by an order of magnitude or more.

The issue of dealing with differences between the training and test data affects any machine-learning based approach, including the statistical learning technique (*boosted regression trees* [21]) underlying the models proposed in this paper, meaning that we need to account for such cases explicitly. To first illustrate the problem, we trained a regression tree model to estimate the CPU overhead of the *Scan* operators in a TPC-H workload, executed on small (scale-factor: 1-4) TPC-H databases and then deployed the model to estimate CPU overhead for scans executed on larger (scale-factor: 6-10) data sets. The results are shown in Figure 3 – as soon as scans access larger data sets, the model systematically underestimates resource usage.

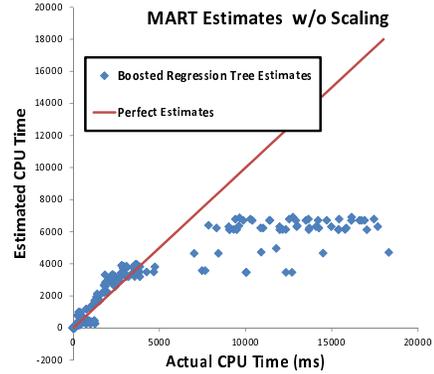

**Figure 3:** Boosted regression trees fit queries that are similar to the training data extremely well, but do not generalize well beyond the training data.

Given the importance of robustness for any component used at the core of query processing, such failure cases may negate any estimation improvements seen for the queries that are similar to the training instances. As a consequence, the problem we address in this paper is how to engineer a resource-estimation framework which (i) uses statistical models to provide estimates that are more accurate than the current state-of-the-art and (ii) at the same time are *robust* in that they generalize beyond the training data.

**Key Ideas:** One of the key decisions when using statistical models for resource estimation is the choice of the underlying machine learning technique itself. When provided with sufficient training data, models such as regression trees or the KCCA approach of [15] are able to fit very complex distributions of the training data with high accuracy. However, as illustrated above, they do not "extrapolate" beyond the training examples they were given. In contrast, much simpler models with a fixed functional form "naturally" extrapolate beyond the training data – for example, a simple linear model can directly encode the fact that CPU usage in a filter scales linearly with the number of input tuples. However, such models are generally limited in the set of functions they can fit well. For example, linear models fail to accurately model dependencies between features and resource usage that are more complex than the simple linear dependency in the example above. Moreover, the choice of the "correct" scaling function for different operators and features may vary significantly: for example, we expect the CPU-usage of a sort operator to scale proportionally to $n \log n$ (where $n$ is the input size), and not linearly.

Our approach overcomes these issues by combining both types of models. We use *boosted regression trees* [21] as the baseline statistical models for areas in the feature space that are "well-covered" by the training examples. When we encounter "outlier" feature values that are very different from training examples, we combine them with additional so-called *scaling functions* to extrapolate beyond the data values seen in training. These scaling functions explicitly model the asymptotic behavior when changing the "outlier" feature value in question; we select their functional form with a different training procedure that systematically varies feature values over a wide range and evaluates the ability of different scaling functions to fit the resulting resource usage curves.

## 1.2 Overview of the Proposed Approach

The technique we describe in this paper uses machine learning to construct models of resource usage based on *training examples* obtained by executing a set of SQL query plans and measuring their resource consumption. We construct separate models for each type of resource; in this paper we concentrate on two types of resources: the CPU time taken by a query and the number of logical I/O operation it performs. To construct the models, each query $\mathcal{Q}$ in the training data is modeled in the form of various *features*, which encode all characteristics of $\mathcal{Q}$ important for the estimation task. After training, each model takes a set of features describing a query as input and outputs an estimate of how much of a specific resource this query will use.

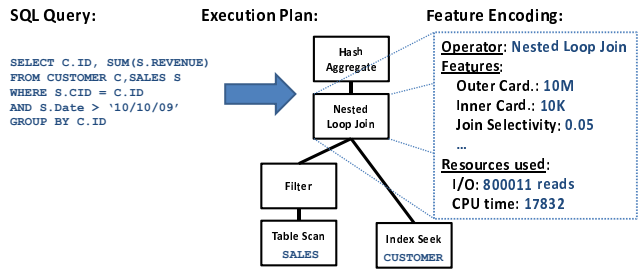

**Figure 4:** Generating Features for a SQL Query

**Features:** The features we use for this task model a query on the level of individual operators (see Figure 4); this has the advantages of (a) allowing us to generalize to query plans not seen during training by combining the corresponding operators and (b) estimate resource usage not only at the granularity of an entire query but also at the level of an operator or *operator pipeline* (i.e., a maximal subtree of concurrently executing operators in a query plan). This is important for applications such as scheduling (since pipelines that do not execute concurrently do not compete for resources). The features themselves are obtained from the query execution plan and the database metadata, meaning that they can be obtained at very low overhead before executing a query.

**Learning Model:** As the underlying machine learning approach we use a variant of (boosted) *regression trees* [21]. Because these models are able to arbitrarily break the domain of input features and does not rely on any specific functional form, they are able to fit even complex dependencies between input features and resource usage with high accuracy. In addition, in cases of "outlier" features much different from the observed training data, we combine the models with scaling functions that model the asymptotic behavior of changes in these features using a fixed functional form. Since the resulting combined model incorporates both regression trees and scaling functions, the resulting function can be much more complex than the scaling function itself (i.e., a linear scaling function in a single feature does *not* imply that the resulting prediction varies linearly with this feature over all feature values), since the regression tree can adapt its estimate to "correct" for the effects of the scaling function within the range of feature values seen during training.

**Contributions:** Our paper makes the following contributions:
(1) We propose a hybrid model that combines the ability to accurately fit complex dependencies between input features and the resources usage with fixed-form scaling functions which are selected to model asymptotic operator behavior.
(2) Using a number of real-life and synthetic workloads and data sets, we show that the proposed technique is significantly more accurate than competing approaches, such as the optimizer cost model itself, various other models build on the same set of features used by our approach (including linear and SVM regression, combined with various Kernel functions), the *operator-level* model of [8], and a combination of decision trees and regression designed to approximate *transform regression* [18]. The evaluation is based on Microsoft SQL Server and large scale real-life and benchmark data sets and workloads.
(3) We conduct an extensive evaluation of our technique's ability to generalize in scenarios where the training data used to build the model is very different from the queries it is applied to. In particular, we systematically vary parameters such as data size and evaluate our approach in numerous scenarios where training and test queries are completely different, i.e., they do not share data, schemata or query templates.
(4) Finally, we show that the overhead of using the resulting models is very small, making them suitable for use in the applications outlined previously.

**Constraints:** In this paper, we focus on resource estimation of CPU time and logical I/Os for a query execution plan. These resources are primarily a function of the execution plan itself and do not depend on transient effects such as which other queries are concurrently executing. Estimating other resources such as *physical* I/O, which can depend heavily on runtime effects is important but beyond the scope of this paper.

## 2. RELATED WORK

Recent work has explored the use of machine-learning based techniques for the estimation of both run-times as well as resource usage of SQL queries, both for queries in isolation [15, 8], as well as in the context of interactions between concurrently executing queries [6, 12]. One key limitation of the approach of [15] is that the resource estimate for a query $\mathcal{Q}$ is obtained by *averaging* the resource characteristics of the three queries in the training data that are the most similar to $\mathcal{Q}$ after mapping $\mathcal{Q}$'s feature vector using *Kernel Canonical Correlation Analysis* (KCCA) into a suitable similarity space. This means that e.g., when attempting to estimate the resource consumption of a query which is significantly more expensive than all training examples, the estimated resource value can never be larger than the ones encountered in the training queries. Thus, this technique is not capable of "extrapolating" beyond the training data. Also, this approach models a SQL query at the level of a plan template, using only the number of instances of each physical database operator type and – for each operator type – the aggregate cardinality as features. This makes the approach vulnerable to changes between training and test data not encoded in this feature set. For example, when the training and test queries use different schemas and databases, the estimated run-times were up to multiple orders of magnitude longer than the actual time the queries ran (see Experiment 4, Figure 15 in [15]).

The approach of [8] side-steps some of these issues by offering both *plan-template models* as well as *operator-level models* for the execution time of queries, as well as a *hybrid model* integrating these two. While this paper does not model resource consumption explicitly, the proposed feature sets and models could conceivably be used for this task. We will concentrate on the proposed operator-level models in the following (and compare our approach to them in the experiments), as it – unlike the template-based approach – has (some of) the generalization properties we are inter-ested in. One main limitation of this model is the machine-learning approach chosen: because the authors use *linear* regression models for each operator, this means that they implicitly force the output of the model to vary linearly with each input feature. Unfortunately, for many features the relationship between resource usage and feature value is more complex and non-linear, resulting in inaccurate estimates. In contrast, the approach proposed in this paper is – due to the use of boosted regression trees as a basis model – able to fit non-linear dependencies found in the training data. Moreover, when modeling the effects of a feature value significantly larger or smaller than seen during training, our approach is able to use various, suitably chosen functional forms to "scale up" the effects of this particular feature. Note that the experimental evaluation in [8] restricts itself to queries from the TPC-H benchmark and only does a single experiment where training and test workload are not from the same query distribution. As a consequence, it is not clear if the techniques proposed in [8] generalize to changes in workloads.

The work of [22] explores a similar approach in the context of XML cost estimation. Similar to our approach, [22] uses operator-level models of cost, but the underlying machine learning model is based on *transform regression*, which corresponds to a combination of piece-wise linear models. We will evaluate a similar piece-wise regression approach in our experiments.

One other related paper in the work of [6], which uses statistical techniques to predict the completion times of batch query workloads when multiple queries are running concurrently. This technique also works by constructing a model based on example executions; however, in this work queries are modeled through a limited set of *query types* meaning that the model needs to observe every query type a priori and can not generalize to new "ad-hoc" queries. The same holds for the approach of [12], which assumes complete knowledge of the space of *all* queries (and all execution plans) the model is used for later as part of the training.

Finally, our technique is also related to various approaches used in the context of cardinality estimation (such as LEO [19], or self-tuning histogram techniques [5, 9], etc.) which use execution feedback as "training data" to improve cardinality estimation inside the query optimizer. One difference to our work is that our technique aims to estimate resource consumption, which is a function of a number of different factors, only one of which is cardinality.

Model-based estimates of cost and resource consumption are also used in the context of optimization of MapReduce jobs [17] and server consolidation [10]. Both of these approaches operate at a very different granularity than the approaches of this paper and it is not clear how to apply them to our scenario.

## 3. OUR APPROACH

In this section, we will give an overview of our approach, which consists of two distinct phases: the off-line model training and the on-line estimation (see Figure 5). During the training phase, we construct a number of (regression tree) models: for each type of physical database operator $o$ (e.g., *Table Scan*, *Merge Join*, *Sort*,...) and resource type, we train a *default model* $\mathcal{DM}_o$ that is used to estimate resource usage for this operator/resource. This model is invoked when the input query is similar to the training examples. In addition, we use a separate training phase that selects the appropriate scaling function for different resource/operator combinations. We then form *combined models* consisting of scaling functions and different regression tree models. The models are retained, but unlike [15] we do not need to retain the training examples themselves. Since there is only a small number of database operators (and a small number of combined models for each), the space requirements for our approach are not excessive.

When deploying the model to estimate resource usage for a query $\mathcal{Q}$, we first compute the feature values for all operators in $\mathcal{Q}$'s execution plan. As we will show when describing the features in detail, the feature values can be derived with virtually no overhead from the execution plan of the query and metadata of the queried tables/indexes. We then use the feature values to select the appropriate model for each operator, and use it to compute estimates.

In the following, we will describe the individual components of our approach: first (Section 4), we describe the properties of *Multiple Additive Regression-Trees* [21], which we use as the underlying machine-learning model. In Section 5 we then describe how we model a SQL query plan. Finally, we describe which combined models we train initially and the *Model Selection* technique we use to select among them online in Section 6.

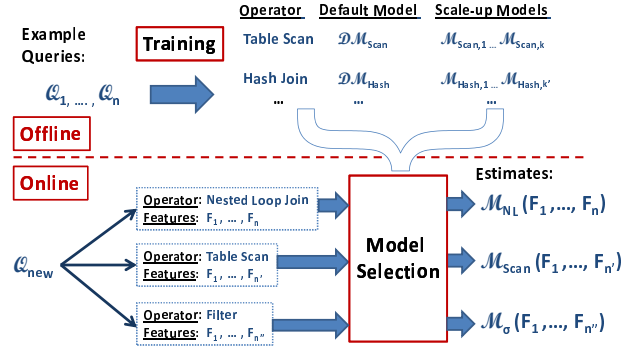

**Figure 5:** Overview over Training and Deployment

## 4. USING REGRESSION TREES TO ESTIMATE RESOURCE CONSUMPTION

The learning method we use to compute estimates of the resource consumption is based on *Multiple Additive Regression-Trees* (*MART*), which are in turn based on the *Stochastic Gradient Boosting* paradigm described in [14]. In the following, we will give a brief high-level overview of the technique and introduce its relevant properties; a more detailed description can be found in [21].

MART models training data by iteratively building a sequence of regression trees that recursively partition the space defined by the set of features using binary split operations. A regression tree is a binary decision tree, where each internal node splits the features space into two by comparing the value of a chosen feature with a pre-computed threshold; once a terminal node is reached, an optimal regression value is returned for all the data falling into the region. At the end of each iteration, the estimation errors for the training data are computed and the regression tree in the next iteration is constructed to fit this residual error. The resulting model is then added to the existing ones, compensating for the errors seen in the prior model and the combined model used to compute the residual errors for the next iteration.

**Properties of MART:** MART has three important properties that are crucial for our scenario. First, by its internal use of regression trees, which are able to "break" the domain of each feature arbitrarily, it is able to handle the non-linear dependencies between the feature values and the required resources without using explicit *binning* as a preprocessing step. As a result, MART is able to fit complex resource-consumption curves that can not easily be expressed using simpler models with a fixed functional form (e.g., linear regression models). Second, MART is able to handle diverse sets of features as it does not require transformations to normalize the inputs into zero mean and unit variance which is essential for other algorithms such as *logistic regression* or *neural nets*. Finally, the functions resulting from the regression trees need not be continuous, which is important for some database operators (e.g., multi-pass sorts, for which resource consumptions may 'jump' with an increasing number of passes).

## 5. MODELING RESOURCE USE

In this section, we will describe the features that we use as the input to the various models used to estimate resource consumption. For this purpose, we first need to specify the granularity at which we model the incoming queries: while most previous work uses features defined at the level of a query (plan) template, i.e., the features describe properties of the entire query (plan), our work uses a more fine-grained approach which defines features and models at the level of individual operators. In the next two sections we first discuss the properties of both approaches before introducing the details of the features we use in Section 5.3.

### 5.1 Template-based Models

In cases where the distributions of test and training queries are very similar (for example, if they are all chosen from the same closed set of query templates, with limited variation in the estimated quantity within a single template), then only a small number of "high-level" features often suffice, as these allow the model to identify the template in question and "use" similar training examples from the same template to estimate the resource usage or run-time. Models of this type are the model described in [15], the *plan-level* model of [8] as well as the model of [12]. The main advantage of these approaches is that – if there is little change in resource usage within a template – they requires little training data, which, given that each training data point results from executing a potentially expensive SQL query, is a valid concern.

However, the major limitation of modeling queries at a template level is that it cannot account for changes within a template not captured by one of these "high-level" features. To illustrate this, consider the features introduced in [15], which use the execution plan $P$ of a query to generate the feature vector $\mathcal{V}$ as follows: for each physical operator supported by the DBMS, $\mathcal{V}$ contains two values: (a) the number of occurrences of the operator in $P$ and (b) the sum of the cardinalities of all instances of this operator. As a result of this modeling, their model implicitly requires that the query plans from the training and test sets are identical with respect to all properties that are not captured by these features. For example, consider the simple case of a query plan consisting only of a *table scan*, followed by a *filter* operator. Now, if a model is trained using examples of this plan scanning a table $T_{train}$, but is used to estimate resources for the same query plan applied to a very different table $T_{test}$, there is no way for the model to detect these differences. For example, if the columns in $T_{test}$ are more than $10\times$ wider than the ones in $T_{train}$, the model would not be able to capture the resulting increase in resource usage or run-time, leading to estimates that are "off" by an order of magnitude.

### 5.2 Operator-level Models

Given the importance of generalizing to "ad-hoc" queries not seen during training, we need to consider modeling queries at a finer granularity and include any properties we expect to vary between the training examples and the queries we use the model for in the feature set. For this purpose, we use the fact that SQL queries are broken down into physical operators within SQL execution engines already (similarly to [8] and [22]) and train different models for each type of SQL operator.

Because this approach mimics the way SQL queries are composed, this (as was also pointed out in [8]) allows us to generalize to previously unseen query plans trivially by combining the models corresponding to the operators in the plan. This also allows us to make predictions at a finer granularity than the full query, which is particularly important in the context of scheduling: for complex queries containing multiple blocking operators, the minimal unit of concurrently executing operators is not the entire query plan, but a *pipeline*; hence, the resource requirements of two pipelines that do not execute concurrently are not incurred by the system at the same time. Note that there are some special cases such as build/probe phases of hash joins (see [16]) for which we may want to break down resource consumption at even finer granularity.

### 5.3 Features

In this section we describe the features we use as input to the operator-level models. These are divided into two categories: *global features* which are relevant for all operators and *operator-specific features* which are relevant for specific (sets of) operators only.

A list of the global and operator-specific features that we use in our experiments can be found in Table 1 and Table 2, respectively. All of these features are numeric, with the exception of OUTPUTUSAGE, which encodes the operator type of the parent operator as an integer. Note that some of the features require knowledge of the number of tuples or bytes in the input/output of an operator, which are typically not known exactly until a query has completed. In these cases, we use the optimizer estimates in place of the true values; the only exception to this are operators that scan an entire table for which we can determine the exact counts a priori.

However, in our experimental evaluation we will perform two sets of experiments – one based on exact values of these input features and one based on optimizer-estimates. This will allow us to differentiate between the error incurred as a result of the way we model resource consumption itself and the error incurred as a result of inexact cardinality estimates.

| Name | Description | Notes |
|---|---|---|
| COUT | # of output tuples | |
| SOUTAVG | Avg. width of output tuples | |
| SOUTTOT | Total Number of bytes output | |
| CIN | # of input tuples | 1 feature per child |
| SINAVG | Avg. width of input tuples | 1 feature per child |
| SINTOT | Total number of bytes input | 1 feature per child |
| OUTPUTUSAGE | Type of parent operator | Categorical Feature |

**Table 1: Global Features used for all Operators**

| Name | Description | Operator |
|---|---|---|
| TSIZE | Size of input table in tuples | Seek/Scan |
| PAGES | Size of input table in pages | Seek/Scan |
| TCOLUMNS | Number of columns in a tuple | Seek/Scan |
| ESTIOCOST | Optimizer-estimated I/O cost | Seek/Scan |
| INDEXDEPTH | # Levels of Index in access path | Seek |
| HASHOPAVG | # Hashing operations per tuple | Hash Agg./Join |
| HASHOPTOT | HASHOPAVG × # Tuples | Hash Agg./Join |
| CHASHCOL | # columns involved in Hash | Hash Agg. |
| CINNERCOL | # columns involved in Join (Inner) | Joins |
| COUTERCOL | # columns involved in Join (Outer) | Joins |
| SSEEKTABLE | # Tuples in inner table | Nested Loop |
| MINCOMP | # Tuples × sort columns | Sort |
| CSORTCOL | # columns involved in Sort | Sort |
| SINSUM | Tot. bytes input in all children | Merge Join |

**Table 2: Operator-specific Features**

**Extensions:** Note that the proposed feature set should not be interpreted as complete; there are a number of important properties

(which may differ between test and training data) that are not captured. Instead, this paper gives a framework for resource estimation, which – given that our models can handle non-linear dependencies and do not require any form of feature normalization – can be extended relatively easily by a database engine developer with additional features.

Expanding the feature set to include such factors as the complexity of filter predicates (which impact CPU time as well as (indirectly) the accuracy of cardinality estimation), improved modeling of additional I/O requests and CPU time caused by disk spills, estimates of the distinct number of values in a join/merge, etc., is virtually certain to improve estimation accuracy even further. Also, additional features would likely be required to account for differences in operators and query processing on different database engines than the one (MS SQL Server) used in our evaluation.

## 6. MODEL SELECTION AND TRAINING

In this section we will describe (a) the characteristics of the combinations of models and scaling functions our approach trains offline and (b) the online method we use to select among them for a given input query. As described in the introduction, our approach extrapolates beyond the training examples seen by using a combination of a MART model and an appropriately chosen scaling function to estimate resource consumption. For example, we may use a "default" MART model $\mathcal{M}$, which takes in a number of features, including the number of input tuples, to predict the CPU use of a *Filter* operator. However, in cases where this number of input tuples is much larger/smaller than the examples seen during training, we switch to a different *combined model* which consists of a different MART model $\mathcal{M}'$, which predicts the average CPU usage for *each* input tuple (something that we can still assess using the existing training data), and a *scaling function* to scale up this estimate, which in this example means multiplying the per-tuple CPU usage by the total number of input tuples (since we expect the CPU usage in a *Filter* node to scale linearly with the input size).

This approach gives us the flexibility to chose a different scaling function for different operator/feature combinations. We use a separate training process to select the appropriate scaling functions, which we will describe later in this section.

**Notation:** A MART model $\mathcal{M}$ is specified via (i) the features $\mathcal{F} = \{F_1, \ldots, F_k\}$ it uses to predict resource consumption and (ii) the resource $\mathcal{R}$ it predicts; we write this as

$$\mathcal{M}(F_1, \ldots, F_k) \to \mathcal{R}.$$

In some cases the scaled models are not trained on the feature values themselves, but on functions $g(F_i)$ of the feature values, in which case we use the notation $\mathcal{M}(g(F_i), \ldots, F_k)$. For simplicity, we will use the same notation $F_i$ when referring to either the *feature name* or the *value* of the feature, as the context should make it clear what is meant. We use the notation $val(F_i) = \{v_1, \ldots v_l\}$ to denote the set of values for the feature $F_i$ seen during training; we use $low(F_i) = \min_{j=1,\ldots,l} v_j$ and $high(F_i) = \max_{j=1,\ldots,l} v_j$ to denote the smallest/largest value in this set.

Now, consider the example described above: our default MART model $\mathcal{M}$ estimates the CPU time in a filter operator as a function of the following features: (1) the number of input tuples (CIN), (2) row width (SINAVG) and (3) the number of output tuples (COUT). We write this as

$$\mathcal{M}(\text{CIN}, \text{SINAVG}, \text{COUT}) \to CPU.$$

However, when faced with a query containing a filter for which the feature CIN had a value much larger than all training instances, we instead use a model $\mathcal{M}'$, which uses the same set of features (with the exception of the feature used for scaling), but instead of estimating the total CPU time, estimates the CPU time per single input tuple. We write this as

$$\mathcal{M}'(\text{SINAVG}, \text{COUT}) \to \frac{CPU}{\text{CIN}},$$

and refer to $\mathcal{M}'$ as the *scaled model* (since we scale its output estimate) and $\mathcal{M}$ as the *original model*. Now, we use a *scaling function* $SCALE_{\text{CPU,CIN}}()$ to scale up the estimate produced by $\mathcal{M}'$ to account for the value of CIN. In this example, we use *linear* scaling, i.e., we multiply the initial estimate by the number of tuples; here, we refer to the resulting overall function as the *combined model*:

$$\begin{aligned} CPU &= SCALE_{\text{CPU,CIN}}(\mathcal{M}'(\text{SINAVG}, \text{COUT})) \\ &= \underbrace{\text{CIN} \times}_{\text{Scaling function}} \underbrace{\mathcal{M}'(\text{SINAVG}, \text{COUT})}_{\text{Scaled Model}}. \end{aligned}$$

When the used scaling functions contain a feature $F_i$, we also say that we "scale the estimate by $F_i$".

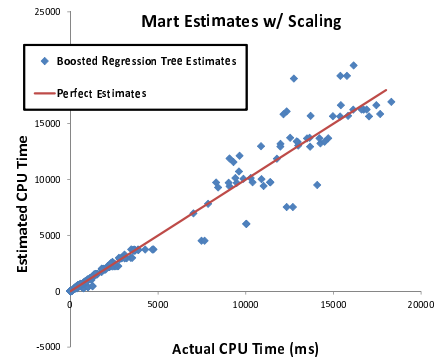

**Figure 6: When combining MART and Scaling, accuracy improves significantly for feature value not seen in training.**

To illustrate the expressive power of combining MART models with scaling, we re-ran the experiment depicted earlier in Figure 3 in Section 1, using the combination of MART and a linear scaling function for *all* estimates. Recall that in this experiment, the training data included only queries executed on smaller datasets, whereas the model was tested on queries consuming comparatively more data. The resulting estimation accuracy is shown in Figure 6. As we can see, the resulting model retains the extremely accurate modeling of the original MART model for the "small" queries in the lower left corner, and at the same time generalizes much better to queries with much larger CPU times.

### 6.1 Defining the Combined Models

In the following, we will define the appropriate combined model for an operator and a given set of features that are "outliers" (i.e., either significantly larger or smaller than the feature values seen during training). We will first consider the case of only a single outlier feature $\hat{F}_i$ and then extend our approach to multiple such features. As we illustrated with the example of Section 6, a combined model is defined by the (i) appropriate scaling function $g(\hat{F}_i)$ and (ii) the *scaled model*, which corresponds to a modified version of the operator's default model. Note that different scaling functions may be used to account for different "outlier" features (even for a single operator). The choice of the scaling function is based on a set of experiments where we systematically vary feature values and observe the resulting changes to resource consumption, which

1560

we will explain in detail in Section 6.2. The scaled model is identical to the default model in terms of features set and training data, with the exception of 3 modifications:
(1) The scaled model does not estimate the total resource consumption, but instead the resource consumption for a single unit of $\hat{F}_i$. Consider the earlier example, where $\mathcal{M}'$ did not estimate the *total* CPU time, but the CPU time *per input tuple* (which is then scaled up). In general, when training a scaled model $\mathcal{M}'$, which we want to use in combination with a scaling function $g(\hat{F}_i)$, we modify the training data used such that – for each training example – the value of resource usage to be predicted is divided by $g(\hat{F}_i)$.
(2) $\hat{F}_i$ is removed from the list of input features to $\mathcal{M}'$.
(3) We also need to consider the effect of scaling when there are dependencies between $\hat{F}_i$ and other input features of $\mathcal{M}'$. Again, we use an example: consider the case of a default model

$$\mathcal{M}(\text{CIN}, \text{SINTOT}) \to \mathcal{R}$$

and a query that contains 3× as many input tuples as any query in the training set. Because both features (CIN, SINTOT) are a function of the number of input tuples, both of them therefore are likely take on values that are much larger than the ones in the training data. However, when we now scale using $g(\text{CIN})$, we're implicitly constructing a model for the resource consumption of a single input tuple and correspondingly should reduce the feature value of SINTOT (which corresponds to the product of the number of tuples and the average row width) to the (average) number of bytes in a single output tuple. Otherwise, SINTOT is likely to continue to be an outlier value, causing us to scale multiple times due to a single cause – an excessive number of input tuples. Thus, we now divide (when training the model) the values of SINTOT by the value of the outlier feature (CIN), which results in the model

$$\mathcal{M}'(\frac{\text{SINTOT}}{\text{CIN}}) \to \frac{\mathcal{R}}{\text{CIN}}.$$

This means that we need to do the appropriate "normalization" of dependent features both when training $\mathcal{M}'$ as well as when using the combined model to predict resource usage.

In general, we perform normalization only for the values of features that are dependent on the outlier feature; here, we define dependence as meaning that a change in the value of the outlier implies a change in the value of the dependent feature. For example, CIN and SINTOT are dependent, whereas CIN and SINAVG are not. We have summarized all feature-pairs that we consider dependent for the purpose of feature normalization in table 3 – dependent features are marked using the symbol ■.
**Scaling by Multiple Features:** In the cases where we want to scale by more than one feature (e.g., if the number of tuples and the column widths are both outliers), we construct the corresponding model by first scaling by the one feature, and then repeating the scaling construction for the resulting scaled model $\mathcal{M}'$ for the next feature, etc. To illustrate this, consider the original model

$$\mathcal{M}(\text{TSIZE}, \text{SOUTAVG}, \text{TCOLUMNS}) \to CPU,$$

which estimates the CPU time of a *Index Seek* operator. First, consider the case that TSIZE is much larger than the training examples; in this example, we use logarithmic scaling (assuming that index seek CPU is proportional to index depth, which grows logarithmically with table size). The resulting combined model has the form

$$\log_2(\text{TSIZE}) \times \mathcal{M}'(\text{SOUTAVG}, \text{TCOLUMNS}).$$

If we then scale $\mathcal{M}'$ by SOUTAVG, the resulting combined model becomes

$$\log_2(\text{TSIZE}) \times \text{SOUTAVG} \times \mathcal{M}''(\text{TCOLUMNS}).$$

Table 3: Feature Dependencies

While our framework supports scaling by arbitrary numbers of features, we only use at most two of them in our experiments. This keeps the number of models that we need to store manageable.
**Selecting the Default Models** $\mathcal{DM}_o$: Finally, we select – among all trained models – the default model for each operator type $o$ as the one that gives the minimum estimation error for the set of training queries. Note that this means that the default model may already incorporate scaling.

## 6.2 Scaling Functions

In this section we describe the framework used to select the scaling function for a given operator and feature $F_i$. While in some cases (such as the example of the *Filter* operator) the appropriate scaling function is apparent from our understanding of SQL query processing, this does not hold for all operators. For these instances, we set up experiments in which we synthetically generated a large set of SQL queries containing the operator in question for which the value of the feature in question varies over a wide range. Note that these queries were designed in such a way that the value of all features without a dependency on $F_i$ were kept constant, and (because of the way we handle dependent features) for all features $F'$ with a dependency, the ratio $\frac{F'}{F_i}$ was kept constant. We execute these queries, recording their resource consumption and then use this data to select the appropriate scaling function by fitting the resulting feature/resource curves with different (simple) models with fixed functional forms. Among these, we then select the function that models the data the most accurately (based on the $L_2$ error).

To illustrate this framework in more detail, we use the example of modeling the CPU consumption of *Sort* operators when scaling the number of input tuples CIN; here, we generate observations using the SQL template

**SELECT * FROM lineitem WHERE l_orderkey $\leq$ [$t_1$] ORDER BY Random_Funtion(),**

where we varied the value of $t_1$ and used random numbers as the sort key to ensure that the sort order of the tuples is not correlated to their order on disk. We then attempted to fit the resource usage curve using a number of different functions, including linear scaling (where we use a function CPU = $\alpha \times$ CIN and fit the value of $\alpha$ using least-squares regression), $n \log n$-scaling (i.e., using CPU $= \alpha \times \text{CIN} \times \log_2(\text{CIN})$), exponential scaling (CPU $= \alpha \times \text{CIN}^\beta$ for different values of $\beta$), and logarithmic scaling (CPU $= \alpha \times \log_2(\text{CIN})$). We show the observations and resulting curves for $n \log n$-scaling and quadratic scaling in Figure 7; note that the CPU usage values are the same in both graphs, but that we change the x-axis, giving us different curves. As one would expect, in this case the $n \log n$-scaling function performed best when fitting the data.
**Multi-feature Scaling:** For join operators, which have two inputs, we also consider scaling functions that have two input variables; for example, when scaling *Merge Joins*, which have two inputs that are treated symmetrically, a scaling function that increases proportionally to the sum of the input sizes becomes a candidate. For joins, we also consider other scaling functions that have two input parameters, such as scaling by the product of both features, scaling by the product of one feature and the logarithm of the 2nd one, etc. To illustrate how the observed data is fit by the resulting fitted curves, we plotted the observed and predicted CPU consumption for *Index Nested Loop Joins* in Figure 8.
**Non-scaling Features:** Finally, when estimating the I/O operations of a query plan, there are a number of features that are either irrelevant or only model second-order effects and hence are never considered for scaling up. These features are HASHOPAVG, HASHOP-TOT, CHASHCOL, CINNERCOL, COUTERCOL, MINComp and CSORTCOL. Obviously, non-numeric features such as OUTPU-TUSAGE are never candidates scaling.

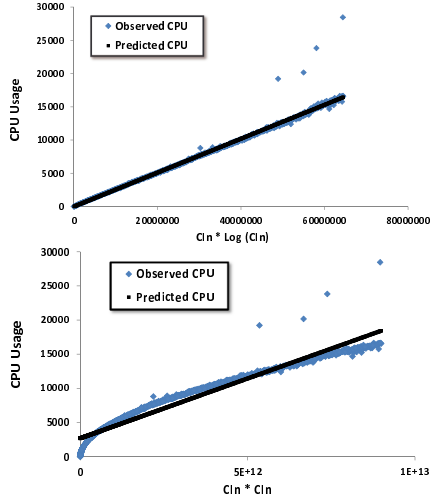

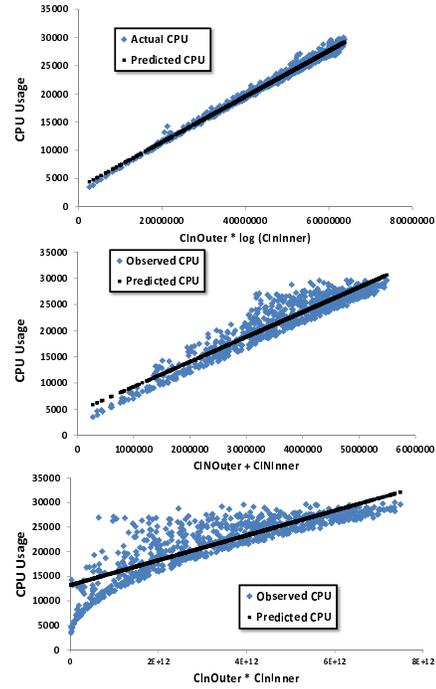

**Figure 7:** Evaluating different scaling functions for the CPU-consumption of sort operators: scaling by $\text{CIN} \times \log_2(\text{CIN})$ fits the data very with high accuracy.

**Figure 8:** Evaluating different scaling functions for the CPU-consumption of index nested loop joins: scaling by $\text{CINOUTER} \times \log_2(\text{CININNER})$ fits the data better than the alternative functions.

### 6.3 Model Selection

Given these different models, we now describe how to – for an incoming query $\mathcal{Q}$ – select the model to use among them. This decision is made on the level of each operator in $\mathcal{Q}$'s execution plan. A property we use here is that the relationship between resource usage and all features we consider for scaling is monotonic: a (sufficiently large) increase (or decrease, respectively) in a feature value implies an increase (decrease) of the resource usage as well. Thus, we can use the difference between the value of a given input feature and the largest/smallest values seen for this feature during training as a heuristic quantifying our ability to model the effects of this feature value accurately using a default model. Note that – due to the feature normalization described in Section 6.1, which changes the values a model is trained on – this value may be different for the different candidate models, depending on which feature(s) they scale by.

We first compute, for a feature $F_i$ in the operator we are trying to predict, and each applicable model $\mathcal{M}$, how much it differs from the largest/smallest values seen in training as $out\_ratio(F_i, \mathcal{M})$

$$= \frac{\min \left\{ \max\{low(F_i) - F_i, 0\}, \max\{F_i - high(F_i), 0\} \right\}}{high(F_i) - low(F_i)}.$$

Using this ratio, we then select the model we use to make an estimation as follows: if $out\_ratio(F_i, \mathcal{DM}_o) = 0$ for all features $F_i$ we use the default model $\mathcal{DM}_o$. Otherwise, we use the model that – if we take the maximum of all $out\_ratio$ values over the set of all features – has the smallest maximum. We break any ties by (a) preferring a model using fewer scale-out parameters and (b) if this still results in a tie, using the 2nd-largest $out\_ratio$ to break the tie, etc. Finally, note that while this approach performs well in experiments, it is a simple heuristic that may not perform well when the distribution of feature values within the ranges between their maxima/minima is very uneven. We leave the study of more sophisticated model selection techniques as future work.

## 7. EXPERIMENTAL EVALUATION

In this section we evaluate the accuracy of our techniques as well as various competitors. For this, we need to not only to evaluate the resulting accuracy when we apply models to test queries that are similar to the training set, but also when the test queries are very different; for the latter scenario, we use a large TPC-H workload as the training data and different real-life decision-support and benchmark query workloads as the test set. All experiments were conducted using SQL Server 2010 on a 3.16 GHz Intel Core Duo PC with 8GB of memory.

**Datasets & Workloads:** Overall, we use the following workloads and datasets in our experiments: (1) Over 100 randomly chosen queries from the TPC-DS benchmark [3]. The database size is approx. 10GB. (2) Over 2500 queries from the TPC-H benchmark [3], which we randomly generated using the QGEN tool. We generated the underlying data distribution using the tool published at [2]; here, we chose a high skew-factor of $Z = 2$, meaning that there are very significant differences in the resource consumption between queries (even among queries from the same TPC-H query template). We also varied the overall size of the data, generating data sets corresponding to scale-factors 1, 2, 4, 6, 8, 10. (3) "*Real-1*" is a real-world decision-support and reporting workload over a 9GB sales database. Most of the queries in this workload involve joins of 5-8 tables as well as nested sub-queries. The workload contains 222 distinct queries. (4) "*Real-2*" is a different real-life decision-support workload on a larger data set (12GB) with even more complex queries (with a typical query involving 12 joins). This workload contains a total of 887 queries.

**Alternative Techniques:** We compare the accuracy and generalization ability of the following techniques:

(1) Using the *optimizer estimates* of resource usage – for this purpose, we compute the resource usage for a resource $\mathcal{R}$ as the optimizer-estimate multiplied by an adjustment factor $\alpha_\mathcal{R}$ (which corresponds to the skew of the regression line in Figure 1); we use the training data to set these adjustment factors $\alpha_\mathcal{R}$, such that the $L_2$ error between the observed usage and the estimated usage (after being multiplied by the ratio) is minimized. We compute a different adjustment factor for each operator type. We will refer to this technique as OPT.

(2) The *operator-centric model proposed in [8]*; we use the same set of features proposed in Tables 1 and 2 of this paper and linear regression (combined with feature selection) as the underlying statistical model. We also use the same bottom-up propagation mechanism for resource estimates as in [8], however, instead of propagating (estimated) startup and execution times, we instead propagate the (estimated) cumulative resource usage.[1]

(3) *Pure Linear regression models* for each operator built using the numerical features we proposed in this paper for the same operators (using feature selection to select the most efficient set of features). We will refer to this technique as LINEAR.

(4) The collection of *MART models* for each operator and using the same features, without an explicit scaling component. We will refer to this technique as MART.

(5) A different method to achieve both the ability of model non-linear dependencies between features and predicted resource consumption as well as extrapolation beyond the observed training examples are *Support Vector Machines* (SVM) [20] when combined with suitably chosen Kernel functions. Here, we used the SVM regression package which is part of the WEKA machine learning toolkit [4] and evaluated this approach when combined with all Kernel functions suitable for numeric data which are part of the toolkit (*NormalizedPolyKernel, PolyKernel, Puk,* and *RBFKernel*). For readability, we will only show the result for the best-performing Kernel in each of the following sections (which were the *PolyKernel* in Sections 7.1.1 & 7.1.2 and the *RBFKernel* in Section 7.2).

(6) We also want to compare to the approach of [22] which uses *transform regression* trees [18] as the underlying machine-learning model. Unfortunately, no implementation of transform regression is available to us. Instead, we use the most similar model available to us, which is a modification of MART, which – similar to the approach used in [22] – uses linear regression (in one feature) at each tree node and is a "boosting" approach that generates a sequence of models, where each subsequent model fits the residual error remaining after application of the previous ones (details are to be published in [1]). However, note that some of the remaining details of the two techniques differ. We will refer to this technique as REGTREE.

(7) The combination of *Model Selection and Scaling* described in this paper. We will refer to this technique as SCALING.

**Model Training and Feature Selection:** For all experiments, we trained the underlying MART models (see Section 4) using the $M = 1K$ boosting iterations; each decision tree has at most 10 leaf nodes. As in [8], we use greedy feature selection to determine the best-performing subset of features for each of the different operator-models used by this competitor. When training the SVM based models which have a dependency on one or more training parameters (e.g., the *gamma* parameter used in the *RBFKernel*) we used the methods in WEKA for finding the optimal parameter settings (see: *http://weka.wikispaces.com/Optimizing+parameters*).

## 7.1 Evaluation: Estimation of CPU Time

In this section we evaluate the accuracy of the above techniques for estimating the CPU time used by different queries, as well as the technique's abilities to generalize to queries different from the training data. For this purpose, we conduct three types of experiments: (a) training the underlying models using one workload and then testing the resulting estimation accuracy on a disjoint set of queries from the same workload (b) training the models on a workload and testing on a disjoint set of queries from the same workload which have a different feature distribution, (c) training using one workload and testing on a completely different one (i.e., different schemata, queries, query plans and data), which corresponds to the most challenging scenario for generalization.

In each of these experiments, we track the average relative error, when averaging over all queries in the test set:

$$L_1\_Err = \sum_{\mathcal{Q} \in TestSet} \left| \frac{Estimate_\mathcal{R}(\mathcal{Q}) - True\_Usage_\mathcal{R}(\mathcal{Q})}{Estimate_\mathcal{R}(\mathcal{Q})} \right|.$$

In addition, we also track the faction of queries where the maximum ratio between the estimate and the true value, defined as

$$Ratio\_Err := \max \left\{ \frac{Estimate_\mathcal{R}(\mathcal{Q})}{True\_Usage_\mathcal{R}(\mathcal{Q})}, \frac{True\_Usage_\mathcal{R}(\mathcal{Q})}{Estimate_\mathcal{R}(\mathcal{Q})} \right\}$$

is (a) smaller than 1.5, (b) between 1.5 and 2, and (c) larger than 2. As we will see during the experiments, the relative performance of different techniques with regards to the two error metrics differs in many cases: while the $L_1$ error is sensitive to a small number of queries with large error, the fraction of queries with small ratio error better captures the "typical" performance of various techniques.

Any lack of accuracy for the different models can stem from two sources: one of them being that the models themselves are not accurate, and the other being that – while the model itself is accurate – we cannot quantify the correct values of input features. The latter case happens frequently when factors like the number of input tuples are used as features – since we do not know the exact number of input tuples to an operator before a query has executed, we have to rely on optimizer estimates.

In order to distinguish between the effects of these two sources of error, we ran two sets of experiments: in the first experiment we used the exact cardinalities for any feature that incorporates tuple counts, when available (due to insufficient instrumentation we could not obtain the exact counts for some joins with residual predicates as a side-effect of executing the query). In the second experiment we used optimizer estimates instead. Ultimately, the errors seen in the second experiment are the most important ones, as these are indicative of the performance of different techniques in practice. Note that the competitor based on optimizer cost estimates (OPT) uses estimated cardinalities by default and hence will only be used in that set of experiments.

---

[1] Note that the technique of [8] was proposed for the task of predicting the execution times of queries and not their resource usage itself. However, given the close connection between resource usage and execution cost, we want to evaluate if the proposed model can be used for this task as well.

### 7.1.1 Experiments using Exact Input Features

In the first experiment we conducted in this evaluation we use the set of over 2500 randomly generated TPC-H queries for training and test data; here, we used 80% of the queries for training and the remainder for testing, ensuring that the training and test set did not contain an identical query (i.e., same template and identical parameters). The results are shown in Table 4. As we can see, the proposed scaling approach outperforms all others, both with regards to the $L_1$ error as well as the fraction of queries which are approximated with a small error; only the REGTREE model comes close, followed by the combination of the SVM model + *PolyKernel*. The MART estimator w/o scaling does not perform nearly as well – while it is nearly as good as the SCALING technique in terms of the fraction of queries approximated with a ratio error of less than 1.5 (and much better than the competitors), a small number of outlier queries with large error result in the poor performance in terms of $L_1$ error. Among the other techniques, the operator-level model of [8] outperforms LINEAR.

| Technique | $L_1$_Err | $R \leq 1.5$ | $R \in [1.5, 2]$ | $R > 2$ |
|---|---|---|---|---|
| [8] | 0.34 | 69.17% | 15.81% | 15.02% |
| LINEAR | 0.42 | 60.87% | 24.11% | 15.02% |
| MART | 0.57 | 91.70% | 3.56% | 4.74% |
| SVM(PK) | 0.19 | 84.58% | 10.28% | 5.14% |
| REGTREE | 0.14 | 91.70% | 5.93% | 2.37% |
| SCALING | 0.13 | 94.07% | 1.98% | 3.95% |

Table 4: **Training and Testing on TPC-H (exact features)**

The next experiment uses the same training queries, but – similarly to the example first discussed in Section 1.1 – uses training/test sets that correspond to these queries being executed on different data sizes: in the first experiment, the queries in the training data are executed on smaller databases (i.e., a scale-factor $\leq 4$) and the queries in the test data on large databases (i.e., a scale-factor $\geq 6$) only; in the 2nd experiment, the databases are switched. The results are shown in Table 5. Again, SCALING performs best, with only the SVM model being close in one of the two experiments (and significantly worse in the other). Interestingly, the REGTREE model performs much worse than earlier for this experiment. All other techniques perform much worse.

Finally, to measure the ability to generalize to completely different data sets and query plans, we used the models that we had trained using TPC-H queries to estimate the CPU time for (a) randomly chosen queries executed on the TPC-DS benchmark, (b) the queries from the two different real-life decision support workloads described in the experimental setup. These queries have completely different plans (over different tables and indexes) and – especially in case of the decision support workloads – much larger resource usage. The results are shown in Table 6. Again, SCALING outperforms the other techniques (in many cases quite significantly). The model of [8] performs next-best in terms of $L_1$ error, followed by REGTREE.

### 7.1.2 Experiments using Optimizer Estimates

| Technique | Test Set | $L_1$ | $R \leq 1.5$ | $R \in [1.5, 2]$ | $R > 2$ |
|---|---|---|---|---|---|
| [8] | Large | 1.00 | 50.81% | 18.55% | 30.65% |
| LINEAR | Large | 0.33 | 55.04% | 14.73% | 30.23% |
| MART | Large | 0.37 | 62.79% | 16.28% | 20.93% |
| SVM(PK) | Large | 0.22 | 86.82% | 8.53% | 4.65% |
| REGTREE | Large | 0.36 | 52.71% | 29.46% | 17.83% |
| SCALING | Large | 0.20 | 86.82% | 8.53% | 4.65% |
| [8] | Small | 0.36 | 64.34% | 15.50% | 20.16% |
| LINEAR | Small | 1.09 | 58.87% | 16.13% | 25% |
| MART | Small | 1.94 | 65.32% | 20.16% | 14.52% |
| SVM(PK) | Small | 0.57 | 74.19% | 13.71% | 12.10% |
| REGTREE | Small | 0.39 | 73.39% | 13.71% | 12.90% |
| SCALING | Small | 0.25 | 81.45% | 13.71% | 4.84% |

Table 5: **Training on TPC-H, Testing with different Data Distributions (exact features)**

| Technique | Test Set | $L_1$ | $R \leq 1.5$ | $R \in [1.5, 2]$ | $R > 2$ |
|---|---|---|---|---|---|
| [8] | TPC-DS | 0.71 | 29.76% | 7.14% | 63.10% |
| LINEAR | TPC-DS | 1.47 | 36.90% | 20.24% | 42.86% |
| MART | TPC-DS | 12.30 | 45.24% | 16.67% | 38.10% |
| SVM(PK) | TPC-DS | 1.68 | 46.43% | 22.62% | 30.95% |
| REGTREE | TPC-DS | 0.89 | 34.52% | 28.57% | 36.90% |
| SCALING | TPC-DS | 0.52 | 66.67% | 19.05% | 14.29% |
| [8] | Real-1 | 0.78 | 39.62% | 18.24% | 42.14% |
| LINEAR | Real-1 | 1.00 | 24.53% | 16.35% | 59.12% |
| MART | Real-1 | 16.38 | 60.38% | 15.09% | 24.53% |
| SVM(PK) | Real-1 | 1.09 | 36.48% | 16.98% | 46.54% |
| REGTREE | Real-1 | 1.04 | 55.35% | 15.72% | 28.93% |
| SCALING | Real-1 | 0.62 | 71.07% | 19.50% | 9.43% |
| [8] | Real-2 | 1.13 | 20.73% | 24.51% | 54.77% |
| LINEAR | Real-2 | 1.91 | 27.99% | 15.73% | 56.28% |
| MART | Real-2 | 77.85 | 55.82% | 14.83% | 29.35% |
| SVM(PK) | Real-2 | 5.17 | 20.12% | 15.58% | 64.30% |
| REGTREE | Real-2 | 1.77 | 28.59% | 21.79% | 49.62% |
| SCALING | Real-2 | 0.42 | 66.11% | 20.27% | 13.62% |

Table 6: **Training on TPC-H, Testing on different Workloads/Data (exact features)**

In this section, we repeat all experiments in the previous section, with the main differences being that (a) instead of assuming exact knowledge of input feature values incorporating input/output cardinalities we now use the corresponding optimizer estimates to compute these features and (b) because of this, we can now also compare our approach to the "adjusted" optimizer estimate. The results are shown in Tables 7-9. In addition to modeling resource consumption of operators, this also tests the ability of the different techniques to compensate for errors/bias in the optimizer's cardinality estimation.

| Technique | $L_1$_Err | $R \leq 1.5$ | $R \in [1.5, 2]$ | $R > 2$ |
|---|---|---|---|---|
| OPT | 0.56 | 24.51% | 13.04% | 62.45% |
| [8] | 0.49 | 56.52% | 24.51% | 18.97% |
| LINEAR | 0.46 | 56.52% | 28.85% | 14.62% |
| MART | 0.69 | 81.03% | 12.25% | 6.72% |
| SVM(PK) | 0.29 | 73.52% | 20.55% | 5.93% |
| REGTREE | 0.25 | 82.61% | 11.46% | 5.93% |
| SCALING | 0.26 | 83.00% | 10.67% | 6.32% |

Table 7: **Training and Testing on TPC-H (optimizer-estimated features)**

While the estimation errors increase across the board, as the result of features incorporating errors in cardinality estimates, the overall results are very similar to the ones seen in the previous section, with a few exceptions: while SCALING outperforms all other techniques in nearly all experiments, the technique of [8] is slightly better in terms of $L_1$ error (but not ratio error) for the TPC-DS experiment, and the REGTREE performs very slightly better in the initial experiment; both, however, are significantly worse in the remaining ones. Similarly to the earlier experiments, the advantage of the SCALING technique becomes more pronounced the more the test and training workloads differ, demonstrating the robustness of our technique. Both OPT (mainly due to the adjustment factor $\alpha_{\mathcal{R}}$ not generalizing to different workloads) and MART (due to the inability to scale to outlier feature values) perform poorly for all experiments for which the training and test query workloads are from a different distribution.

| Technique | Test Set | $L_1$ | $R \leq 1.5$ | $R \in [1.5, 2]$ | $R > 2$ |
|---|---|---|---|---|---|
| OPT | Large | 0.59 | 22.48% | 14.73% | 62.79% |
| [8] | Large | 0.59 | 40.31% | 29.46% | 30.23% |
| LINEAR | Large | 0.33 | 57.36% | 16.28% | 26.36% |
| MART | Large | 0.42 | 57.36% | 21.71% | 20.93% |
| SVM(PK) | Large | 0.26 | 76.74% | 18.60% | 15.32% |
| REGTREE | Large | 0.47 | 45.74% | 29.46% | 24.81% |
| SCALING | Large | 0.25 | 80.62% | 13.95% | 5.43% |
| OPT | Small | 0.54 | 26.61% | 11.29% | 62.10% |
| [8] | Small | 1.17 | 49.19% | 17.74% | 33.06% |
| LINEAR | Small | 1.13 | 56.45% | 18.55% | 25% |
| MART | Small | 2.09 | 47.58% | 27.42% | 25% |
| SVM(PK) | Small | 0.64 | 69.35% | 15.32% | 15.32% |
| REGTREE | Small | 0.71 | 53.23% | 20.16% | 26.61% |
| SCALING | Small | 0.41 | 67.74% | 16.94% | 15.32% |

**Table 8:** Training on TPC-H, Testing with different Data Distributions (optimizer-estimated features)

| Technique | Test Set | $L_1$ | $R \leq 1.5$ | $R \in [1.5, 2]$ | $R > 2$ |
|---|---|---|---|---|---|
| OPT | TPC-DS | 2.47 | 23.81% | 5.95% | 70.24% |
| [8] | TPC-DS | 1.29 | 16.67% | 10.71% | 72.62% |
| LINEAR | TPC-DS | 1.40 | 23.81% | 20.24% | 55.95% |
| MART | TPC-DS | 13.12 | 22.62% | 19.05% | 58.33% |
| SVM(PK) | TPC-DS | 1.87 | 19.05% | 29.76% | 51.19% |
| REGTREE | TPC-DS | 4.55 | 21.43% | 22.62% | 55.95% |
| SCALING | TPC-DS | 1.36 | 30.95% | 20.24% | 48.81% |
| OPT | Real-1 | 6.10 | 20.13% | 10.69% | 69.18% |
| [8] | Real-1 | 1.40 | 34.59% | 13.21% | 52.20% |
| LINEAR | Real-1 | 1.17 | 16.35% | 15.09% | 68.55% |
| MART | Real-1 | 16.55 | 28.93% | 16.98% | 54.09% |
| SVM(PK) | Real-1 | 1.18 | 20.13% | 10.06% | 69.81% |
| REGTREE | Real-1 | 2.45 | 18.24% | 8.81% | 72.96% |
| SCALING | Real-1 | 0.83 | 38.36% | 21.38% | 40.25% |
| OPT | Real-2 | 773.40 | 11.20% | 8.93% | 79.88% |
| [8] | Real-2 | 1.52 | 16.34% | 20.42% | 63.24% |
| LINEAR | Real-2 | 2.19 | 20.88% | 13.16% | 65.96% |
| MART | Real-2 | 78.42 | 34.95% | 15.58% | 49.47% |
| SVM(PK) | Real-2 | 7.46 | 13.62% | 10.44% | 75.95% |
| REGTREE | Real-2 | 2.85 | 12.71% | 9.38% | 77.91% |
| SCALING | Real-2 | 1.02 | 42.97% | 19.52% | 37.52% |

**Table 9:** Training on TPC-H, Testing on different Workloads/Data (optimizer-estimated features)

## 7.2 Evaluation: Estimation of I/O

In this section we will evaluate the accuracy of the different techniques for the task of estimating the number of logical I/O operations occurring at an operator. Due to space-constraints, we will only evaluate the performance for the case of using *optimizer-estimated feature values* here; for some operators, the number of logical I/O operations is very closely tied to the optimizer estimates, meaning that for these cases, the results also reflect the model's ability to compensate for any systematic cardinality estimation errors by the optimizer itself.

Here, we only report the results for the four models that performed best. We use the same experimental setup as before and report the results in tables 10-12. Interestingly, the relative performances for three models (LINEAR, [8] and SCALING) is basically identical for all experiments: SCALING performs much better than the competitors, followed by LINEAR and [8], with the latter performing particularly poorly for the generalization experiments using different workloads. In contrast to the earlier approaches, the SVM based models perform significantly worse for the task of I/O prediction than for CPU prediction.

| Technique | $L_1$_Err | $R \leq 1.5$ | $R \in [1.5, 2]$ | $R > 2$ |
|---|---|---|---|---|
| [8] | 0.78 | 73.89% | 9.73% | 16.37% |
| LINEAR | 0.58 | 88.94% | 1.77% | 9.29% |
| SVM(RBF) | 0.69 | 84.50% | 5.04% | 10.47% |
| SCALING | 0.34 | 91.59% | 0.88% | 7.52% |

**Table 10:** Training and Testing on TPC-H (I/O operations)

| Technique | Test Set | $L_1$ | $R \leq 1.5$ | $R \in [1.5, 2]$ | $R > 2$ |
|---|---|---|---|---|---|
| [8] | Large | 0.68 | 68.42% | 10.53% | 21.05% |
| LINEAR | Large | 0.36 | 81.58% | 1.75% | 16.67% |
| SVM(RBF) | Large | 0.17 | 86.82% | 9.30% | 3.89% |
| SCALING | Large | 0.10 | 92.11% | 0.88% | 7.02% |
| [8] | Small | 1.81 | 63.39% | 11.61% | 25% |
| LINEAR | Small | 0.96 | 75.89% | 6.25% | 17.86% |
| SVM(RBF) | Small | 1.65 | 75.97% | 6.20% | 17.83% |
| SCALING | Small | 0.60 | 91.07% | 0.89% | 8.04% |

**Table 11:** Training on TPC-H, Testing with different Data Distributions (I/O operations)

| Technique | Test Set | $L_1$ | $R \leq 1.5$ | $R \in [1.5, 2]$ | $R > 2$ |
|---|---|---|---|---|---|
| [8] | TPC-DS | 3.83 | 24.64% | 15.94% | 59.42% |
| LINEAR | TPC-DS | 1.04 | 43.48% | 14.49% | 42.03% |
| SVM(RBF) | TPC-DS | 11.41 | 13.79% | 8.05% | 78.16% |
| SCALING | TPC-DS | 0.67 | 59.42% | 13.04% | 27.54% |
| [8] | Real-1 | 3.23 | 12.60% | 9.45% | 77.95% |
| LINEAR | Real-1 | 2.31 | 7.09% | 7.87% | 85.04% |
| SVM(RBF) | Real-1 | 8.42 | 15.66% | 7.83% | 76.51% |
| SCALING | Real-1 | 0.66 | 37.80% | 10.24% | 51.97% |
| [8] | Real-2 | 9.72 | 15.01% | 5.52% | 79.46% |
| LINEAR | Real-2 | 5.36 | 29.89% | 4.96% | 65.16% |
| SVM(RBF) | Real-2 | 28.03 | 4.82% | 9.04% | 86.14% |
| SCALING | Real-2 | 0.59 | 71.67% | 7.79% | 20.54% |

**Table 12:** Training on TPC-H, Testing on different Workloads/Data (I/O operations)

In summary, the technique proposed in this paper outperformed all competitors in terms of both accuracy when training and test data are similar as well as generalization performance for practically every single experimental setup and for the prediction of both CPU time as well as logical I/O operations. The differences in accuracy became more pronounced for experiments where test and training data were different.

## 7.3 Evaluation: Training Times and Overhead

In this section we evaluate the training times required for the models we proposed, as well as the overhead of using them. Table 13 shows the training times (in seconds) for various data sizes (which include the time taken for reading in the training data and writing the output model to disk). We used $L = 10$ leaf nodes in

1565

each tree, and $M = 1K$ boosting iterations, which is the same setting we used in the accuracy experiments; note that all models used in the accuracy experiments were trained with less than $5K$ examples. As we can see, the training cost is very small, even for very large training sets. The main overhead for increasing the number of training examples is executing the corresponding queries itself.

|  | Training Examples | | | | | |
| --- | --- | --- | --- | --- | --- | --- |
|  | $5K$ | $10K$ | $20K$ | $40K$ | $80K$ | $160K$ |
| Training Time (s) | 2.61 | 3.55 | 6.86 | 10.99 | 19.62 | 36.75 |

**Table 13: Training Times in seconds when varying the # training examples for $M = 1K$ booting iterations**

**Prediction Cost:** In addition, a crucial issue for the deployment of the resulting models is the overhead of invoking the model at runtime. For this purpose, we measured the overhead for evaluating a MART model for a given input feature set, and obtained an overhead of approximately 0.5 $\mu$s for each call. To put these numbers in perspective, in the context of query optimization, we first measured the time required to optimize queries from each of the different TPC-H query templates on the same hardware. In this experiment, we first optimize all queries, then flushe the plan cache, and measure the time used for optimizing the queries again (thereby making sure that the required statistics and metadata are already in memory). The average optimization time was over 50 ms, meaning that even for several hundreds or even thousands of costing calls, the cost of invoking the MART model for each costing call would not be a significant factor in the overall optimization cost.

**Memory Requirements:** To store the regression tree associated with one boosting iteration, we need to encode the shape of the decision tree (which is encoded in form of the offset of the child nodes) and – for each inner node – the split feature and threshold and – for each leaf node – the resource estimate. Because constrain the trees we use in our experiments to have at most 10 leaf nodes, we can encode node offsets in a single byte (leaf nodes are marked implicitly via 0-offsets). We require one byte to encode the split feature and a 4-byte float for the estimate and split threshold, meaning that a single tree can be encoded in at most 130 bytes. Our experiments use up to 1K boosting iterations, resulting in a model size of up to 127KB. Now, given that the number of distinct SQL operators is small (and for most of them, only a small number of combined models are meaningful), this means that the set of all models can be stored in a few megabytes. Note that these models correspond to an extended cost model for the server, meaning that they only need to be stored once; moreover, their size is independent of the number of training queries or data size.

## 8. CONCLUSION AND FUTURE WORK

In this paper we proposed novel statistical models for resource estimation. Our approach combines superior estimation accuracy with robustness in the face of changes in data size, distribution or novel query plans. We achieve these gains by combining regression tree models with scaling functions, where the latter are used for feature values very different from the ones seen during model training, and the former for all other cases. In addition, by modeling queries at the level of individual operators and utilizing different models for each of them, we can generalize to previously unseen query plans.

We evaluated our approach on Microsoft SQL Server using various large-scale benchmark and real-life decision support workloads. Here our approach outperformed all competing techniques by large margins, and both for experiments based on accurate feature values (reflecting pure modeling accuracy) as well as for experiments based on optimizer-estimates.

While the focus of this work has been on the accuracy of resource estimation itself, the natural follow-up to this work will be to study how to translate the improvements in accuracy to improvements in query processing itself, when deploying our techniques in the context of scheduling, admission control or query optimization. Moreover, it is important to note that the set of features described in the paper are sufficient to cover the various example workloads in our experiments, but are not complete: additional features will be needed to improve the accuracy of the models or cover additional operators, such as user-defined functions.

**Acknowledgements:** We are grateful for the detailed comments from the anonymous reviewers, which have lead to significant improvements in this paper. We would also like to thank the Jim Gray Systems Lab for the support of this work.